\documentclass[11pt,letterpaper]{article}
\usepackage[margin=1in]{geometry}
\usepackage[T1]{fontenc}
\usepackage{lmodern}
\usepackage{amsmath,amssymb,amsthm,mathtools}
\usepackage{booktabs,array,enumitem,microtype}
\usepackage[hidelinks]{hyperref}
\hypersetup{pdftitle={Exact-Quota Facility Location: A Transfer Framework and Three Approximation Guarantees},pdfauthor={Hang Luo}}
\setlist[enumerate]{itemsep=3pt,topsep=5pt}
\setlist[itemize]{itemsep=3pt,topsep=5pt}
\newtheorem{theorem}{Theorem}[section]
\newtheorem{lemma}[theorem]{Lemma}
\newtheorem{corollary}[theorem]{Corollary}

\theoremstyle{remark}

\newcommand{\F}{\mathcal F}
\newcommand{\J}{\mathcal J}
\newcommand{\OPT}{\operatorname{OPT}}
\newcommand{\pos}[1]{\left(#1\right)_+}
\newcommand{\Cp}{C_p}

\newcommand{\Acoef}{\frac{111}{100}}
\newcommand{\Bcoef}{\frac{89}{50}}
\newcommand{\cp}{C_p}

\title{Exact-Quota Facility Location:\\A Transfer Framework and Three Approximation Guarantees}
\author{Hang Luo}
\date{}
\begin{document}
\hypersetup{pageanchor=false}
\begin{titlepage}
\maketitle
\thispagestyle{empty}
\begin{abstract}
\normalsize
Facility location with outliers asks which facilities to open and which clients
to serve, subject to a prescribed service quota. We give a framework for
transferring facility-location cost certificates to this exact-quota setting.
The framework yields deterministic polynomial-time approximation guarantees of
$571/376<1.52$ for metric facility location with outliers, strictly less than
$1.853$ for priority facility location with outliers, and $2.16995<2.17$ for
squared-metric facility location with outliers. Each algorithm serves the
prescribed number of clients and pays the original facility-opening costs.

The framework separates initialization from improvement. A growth process is
stopped when it first reaches the quota; deleting excess clients from its last
batch leaves a payment deficit supported on a bounded number of facilities.
Greedy additions then reoptimize the served client set. A threshold-coverage
representation makes the resulting service cost supermodular and allows the
usual density analysis to compare different client sets. Finally, enumerating
and prepaying a constant number of expensive facilities absorbs the exceptional
costs. A transfer theorem states exactly which initialization certificate is
sufficient and how many prepaid facilities it requires.

The three applications supply this certificate in different ways. The metric
case uses a quota prefix of reassigning JMS; the priority case uses persistent
contributions and a cleanup ordered by service level; and the squared-metric
case uses a local factor-revealing certificate valid for arbitrary client
subsets of the same run. We provide complete finite proofs of the numerical
interfaces, including an exact rational certificate for the squared-metric
bound.
\end{abstract}
\end{titlepage}
\setcounter{page}{1}
\hypersetup{pageanchor=true}

\section{Introduction}
Facility location with outliers asks which facilities to open and which clients
to serve. Opening a facility incurs a cost, and each served client pays its
connection cost to an open facility. Given a quota $p$, the objective is to
minimize these two costs while serving at least $p$ clients. With nonnegative
costs, an optimum can be chosen to serve exactly $p$.

For full-service facility location, a powerful approach controls opening and
connection costs separately, then improves their tradeoff by scaling opening
costs and greedily adding facilities. In the metric setting, Mahdian, Ye, and
Zhang~\cite{MYZ} use this approach to obtain a $1.52$-approximation. With a
prescribed outlier quota, the metric guarantee of JMS~\cite[Section 8.6]{JMS}
is $2$, also recorded as the best known bound by Dabas, Khuller, and
Rivkin~\cite{DKR}. This raises a concrete question: how much of the stronger
opening/connection tradeoff can be retained when the algorithm also chooses
which clients to serve?

Two accounting issues arise. First, an opening event can overshoot the quota:
with a target of $100$ clients, it may increase the number connected from $98$
to $105$. Deleting five new clients restores the quota but removes their
contributions to the facility's opening cost. Second, greedy improvement
changes the best choice of clients: a newly opened facility can make former
outliers cheaper to serve than the current clients. The analysis must compare
the resulting savings with an optimum that may serve a different client set.

Our contribution is a transfer theorem that connects quota initialization to
client-reselecting improvement while retaining separate control of opening
and service costs. Its input is a joint cost bound for an initial solution,
allowing a bounded number of unpaid facilities. Its output is an approximation
guarantee in the original costs, together with an explicit condition on how
many expensive facilities must be guessed and prepaid to absorb the deficit.
The shared proof handles client reselection through supermodularity. The
model-specific task is to construct the initial cost bound. This separation
yields three improved guarantees.

\subsection{Results}
All three models have finite, explicitly given facility candidates,
nonnegative opening costs, and no capacities. In the metric model, connection
costs are distances in a common metric space. In the priority model, each
facility has a service level and each client has a minimum required level;
a connection is allowed when the facility meets that requirement. In the
squared-metric model, the connection matrix satisfies
\[
 \sqrt{c_{ij}}\le\sqrt{c_{ij'}}+\sqrt{c_{i'j'}}+\sqrt{c_{i'j}}
 \qquad(i,i'\in\F,\ j,j'\in\J).
\]
This includes squared Euclidean distances on finite candidate sets.

\begin{theorem}[Three exact-quota guarantees]\label{thm:results}
There are deterministic polynomial-time algorithms for the following problems:
\begin{enumerate}
\item Metric facility location with outliers, with approximation factor
$571/376=1.518617\ldots<1.52$.
\item Priority metric facility location with outliers, with approximation factor
strictly less than $1.853$.
\item Squared-metric facility location with outliers, with approximation factor
$43399/20000=2.16995<2.17$.
\end{enumerate}
Each algorithm connects exactly $p$ clients on feasible instances. Its running
time is polynomial in the binary encoding length of the explicit rational input.
\end{theorem}

\begin{table}[t]
\centering
\begin{tabular}{@{}lcl@{}}
\toprule
Model & Earlier guarantee & This paper\\
\midrule
Metric FLPO & $2$~\cite{JMS} & $571/376<1.52$\\
Priority FLPO & $3$~\cite{LHSW} & $<1.853$\\
Squared-metric FLPO & $9$~\cite{RT} & $2.16995<2.17$\\
\bottomrule
\end{tabular}
\caption{Guarantees for facility location with a prescribed outlier quota.
All rows allow nonuniform opening costs and respect the quota exactly.}
\label{tab:results}
\end{table}

Table~\ref{tab:results} compares these guarantees with the earlier bounds
for the corresponding outlier models. The three applications have distinct
roles in the framework. The metric case shows how to transfer the classical
JMS--MYZ cost tradeoff through a quota prefix. Priority constraints restrict
which connections are legal; they require a different initialization and
cleanup, while the improvement analysis remains the same. Squared-metric
costs keep all connections available but change the local geometric bound;
they require a squared-metric local certificate at the same interface.

\subsection{The transfer mechanism}
For an open set $S$, let $C_p(S)$ be the cost of connecting the cheapest $p$
clients, using only legal connections. The algorithm has three stages:
construct a quota-feasible solution, improve it while reselecting clients,
and restore all opening costs. We describe the certificate that links them.

\paragraph{Keep a joint cost bound after truncation.}
Guess and prepay a constant-size seed set. In the branch containing expensive
facilities of a fixed optimum, let $P$ be their total original cost and put
$R=F^\star-P$. Here $F^\star$ and $C^\star$ are the opening and connection
components of that optimum. After truncation, exempt at most $b$ additional
facilities and let $w$ denote the opening costs with seeds and exceptions
set to zero. For the initial facility set $S_0$ and constants $A,B,t\ge1$,
the required certificate is
\begin{equation}\label{eq:intro-certificate}
 t\,w(S_0)+C_p(S_0)\le AtR+BC^\star.
\end{equation}
The coefficients $A,B$ measure the initializer's losses, and $t$ weights its
opening cost. For a JMS prefix, the retained clients have the $p$ smallest
frozen budgets in a full continuation of the same run. Their total budget is
therefore at most that of the optimum's clients. Applying a local bound to
each optimum facility and its assigned clients gives~\eqref{eq:intro-certificate}.
Deleting only from the last opening event confines the unpaid contributions
to one facility. The priority application obtains the same certificate through
a growth process with a cleanup that preserves legal connections.

\paragraph{Improve the facility set and the served set together.}
For a fixed cost matrix and eligibility relation, $C_p$ is nonincreasing and
supermodular on feasible supersets. In concrete terms, opening a facility
saves less once more facilities are already available, even after the clients
are reselected. This follows from threshold coverage: for any price threshold,
the number of clients reachable below that price is a coverage function;
capping it at $p$ and integrating the remaining deficit gives $C_p$.
Supermodularity makes the individual savings from the optimum's facilities
cover the current connection-cost gap. Greedy additions maximizing savings
per unit of opening cost produce a final set $S_g$ satisfying
\begin{equation}\label{eq:intro-bifactor}
 w(S_g)+C_p(S_g)\le(A+\ln t)R+
 \left(1+\frac{B-1}{t}\right)C^\star.
\end{equation}
This is the step that lets a changing served set inherit an opening/connection
tradeoff.

\paragraph{Restore the unpaid costs.}
When $h$ seeds are the most expensive optimum facilities, pruning more
expensive nonseeds makes each exception cost at most $\tau$, the cheapest
seed cost. Thus $P\ge h\tau$. If both coefficients
in~\eqref{eq:intro-bifactor} are at most $a$, restoring original costs gives
\[
 a(\OPT-P)+P+b\tau
 \le a\OPT+[b-h(a-1)]\tau.
\]
The condition $h(a-1)\ge b$ therefore completes the transfer. It relates the
approximation factor to the number of exceptional facilities and the amount
of enumeration required. Our metric and priority algorithms use two seeds
and one exception; the squared-metric algorithm uses one of each.

\subsection{Related work and the resulting contribution}
\paragraph{Quota prefixes and exceptional costs.}
Charikar, Khuller, Mount, and Narasimhan~\cite[Section 5.3]{CKMN} introduced
expensive-facility guessing and pruning, stopped growth at the quota, removed
excess clients from the final batch, and accounted separately for the last
facility. JMS~\cite[Section 8.6]{JMS} adapted this construction to obtain a
metric $2$-approximation. We retain this treatment of the final event and
connect it to a stronger joint cost bound that can drive subsequent
augmentation. The transfer theorem makes the cost restoration explicit for
$b$ exceptions and $h$ prepaid seeds.

\paragraph{Cost improvement with client reselection.}
JMS~\cite{JMS} supplies the reassigning growth process and arbitrary-star
factor-revealing analysis. MYZ~\cite{MYZ} derives the $(1.11,1.78)$ coefficients
and combines cost scaling with greedy augmentation for its full-service
$1.52$ guarantee. Luo, Han, Shuai, and Wang~\cite{LHSW} also use density-based
additions with client reselection as a heuristic for priority FLPO. Our
analysis connects these improvement steps to the quota initializer: the
supermodularity of $C_p$ compares different served sets, and the joint
certificate yields the two coefficients in~\eqref{eq:intro-bifactor}.
This connection gives a guarantee for the complete initialization-and-improvement
procedure.

\paragraph{Priority connections.}
Luo et al.~\cite{LHSW} give a primal-dual $3$-approximation for priority FLPO.
Their model and cleanup by service level provide the basis for our priority
initializer. We retain frozen clients' contributions, truncate at the quota,
and preserve the separate opening and connection payments through cleanup.
The resulting certificate has $A=1,B=3$ and one exceptional facility.
Combining it with the shared improvement and restoration analysis gives
strictly less than $1.853$.

\paragraph{Squared-metric costs.}
Fernandes, Meira, Miyazawa, and Pedrosa~\cite{FMMP} develop upper-bound
factor-revealing programs and obtain a $2.17$ guarantee for the scaled,
greedily augmented algorithm in the full-service squared-metric model.
We establish the local form needed for quota transfer: a bound on any client
subset of the same run. A finite replication-and-averaging argument connects
this form to their constraint system; an exact rational certificate supplies
our numerical coefficients. The transfer theorem then gives $2.16995$ with
a prescribed outlier quota. In this model, Ren and Tian~\cite{RT} give the
previous $9$-approximation. Zhang, Yuan, and Li~\cite{ZYL} study a multilevel
extension, in which service follows a path through one facility at each level,
and give a $32$-approximation. Our squared-metric result concerns the
single-level model in Table~\ref{tab:results}.

\paragraph{Organization.}
Section~\ref{sec:framework} states and proves the transfer theorem.
Sections~\ref{sec:metric}, \ref{sec:priority}, and \ref{sec:squared}
supply the three initializers and approximation guarantees.
The appendices give the full event-level metric proof, discrete greedy
analysis, and finite local certificates. The numerical proof data and a
solver-free verifier accompany the squared-metric certificate.

\section{A transfer theorem for bounded payment deficits}\label{sec:framework}
Let $\F$ be a finite facility set with nonnegative opening costs $f$.
A feasible open set $S$ has service cost $C(S)\ge0$. We assume that adding
facilities preserves feasibility and that $C$ is polynomial-time computable,
nonincreasing, and supermodular on supersets of a feasible initial set.
In diminishing-savings notation, this means
\[
 C(X)-C(X\cup\{i\})\ge C(Y)-C(Y\cup\{i\})
 \quad\text{for feasible }X\subseteq Y,\ i\notin Y.
\]
The theorem separates two obligations: prove this diminishing-savings property,
and construct a quota-feasible initializer with the joint cost bound below.
For our three models, the first obligation follows from the same coverage argument.

\paragraph{Reselection through threshold coverage.}
For a fixed eligibility relation, define
$d_j(S)=\min\{c_{ij}:i\in S,\ i\text{ eligible for }j\}$, with minimum
$+\infty$ when the set is empty. Let $C_p(S)$ be the sum of the $p$ smallest
$d_j(S)$. A set is feasible precisely when at least $p$ of these values are
finite. The function is computed by scanning eligible edges and sorting.
For $N_u(S)=\{j:d_j(S)\le u\}$,
\begin{equation}\label{eq:framework-coverage}
 C_p(S)=\int_0^\infty\bigl[p-\min\{p,|N_u(S)|\}\bigr]\,du.
\end{equation}
Each $N_u(S)$ is a union of fixed neighborhoods. The marginal gain in capped
coverage decreases when $S$ grows. Integrating gives supermodularity of $C_p$;
on feasible supersets the integral is finite. The proof uses neither triangle
inequalities nor complete eligibility.

\paragraph{Seeds, exceptions, and the certificate.}
Fix integers $h\ge1$ and $b\ge0$. For each $h$-element seed set $H$, retain
\[
 I_H=H\cup\{i:f_i\le\tau\},\qquad
 \tau=\min_{i\in H}f_i,\qquad P=f(H).
\]
An initializer either discards the branch or returns a feasible $S_0\supseteq H$
and $E\subseteq S_0\setminus H$ with $|E|\le b$.
Set $w_i=0$ on $H\cup E$ and $w_i=f_i$ elsewhere, and include every zero-$w$
facility in $S_0$. Starting at $S_0$, repeatedly add an unopened facility
maximizing $(C(S)-C(S\cup\{i\}))/w_i$ whenever this ratio exceeds one.
All unopened facilities have positive $w_i$.
Evaluate every output using $f(S)+C(S)$, and also evaluate every feasible
facility set of size less than $h$.

\begin{theorem}[Bounded-deficit transfer]\label{prop:transfer}\label{app:transfer}
Assume the initializer runs in polynomial time, and that feasibility and
service cost can be evaluated in polynomial time for sets of size less than $h$. Fix $A,B,t\ge1$ and
\[
 a=\max\{A+\ln t,\ 1+(B-1)/t\}>1.
\]
For a fixed optimum $O^\star$ of size at least $h$, suppose its $h$ most
expensive facilities define a nondiscarded branch in which $O^\star$ remains
feasible and the initializer satisfies
\begin{equation}\label{eq:abstract-certificate}
 t\,w(S_0)+C(S_0)\le At(F^\star-P)+BC^\star,
 \qquad F^\star=f(O^\star),\quad C^\star=C(O^\star).
\end{equation}
If $h(a-1)\ge b$, this algorithm is an $a$-approximation, in polynomial time
for fixed $h,b$. The conclusion also holds with any larger $a$ satisfying
these inequalities.
\end{theorem}

\begin{proof}
If $|O^\star|<h$, direct enumeration finds an optimum. Otherwise work in its
good seed branch and put $R=F^\star-P$. Since some reference facilities may
also be exempted, $w(O^\star)\le R$. Supermodularity and monotonicity give
\[
 \sum_{i\in O^\star\setminus S}[C(S)-C(S\cup\{i\})]
 \ge C(S)-C(S\cup O^\star)\ge C(S)-C^\star.
\]
If $D=C(S)-C^\star>R>0$, the best density is at least $D/R>1$.
For a step costing $v$ and reducing the gap to $D'$, therefore
$v\le R(D-D')/D$. Summing steps above $R$ by integration gives a logarithm.
If the first crossing moves from $s>R$ to $d\le R$, combine its opening cost
with the remaining gap:
\[
 v+d\le R+d(1-R/s)\le R+R\ln(s/R).
\]
A crossing exists because there are finitely many additions and greedy cannot
stop above the threshold. Later additions reduce the total accounted cost.
Writing $F_0=w(S_0)$ and $C_0=C(S_0)$ yields
\[
 w(S_g)+C(S_g)\le F_0+C^\star+R+R\ln((C_0-C^\star)/R)
\]
when $C_0>C^\star+R$, and at most $F_0+C_0$ otherwise.
Substitute~\eqref{eq:abstract-certificate} and maximize
$1+\ln x-x/t$ over $x>1$. Its maximum is $\ln t$, attained at $x=t$
(or approached at the endpoint when $t=1$). The other case uses
$1-1/t\le\ln t$. This proves
\begin{equation}\label{eq:framework-final}
 w(S_g)+C(S_g)\le(A+\ln t)R+
 \left(1+\frac{B-1}{t}\right)C^\star\le a(\OPT-P).
\end{equation}
For $R=0$, all reference facilities have zero accounted cost and belong to
$S_0$, so $C_0\le C^\star$. Substitution into $F_0+C_0$ gives the same bound.
Appendix~\ref{sec:greedy} gives the full discrete calculation.

Restoring $H$ and $E$ adds at most $P+b\tau$, while $P\ge h\tau$. Hence
\[
 f(S_g)+C(S_g)\le a\OPT-(a-1)P+b\tau
 \le a\OPT+[b-h(a-1)]\tau\le a\OPT.
\]
The number of seed branches and direct candidates is polynomial for fixed $h$;
there are at most $|\F|$ additions per branch. All choices use computable
costs, not the reference optimum.
\end{proof}

The theorem is an interface for initializers. Its parameters make the
three applications easy to compare:
\begin{center}
\begin{tabular}{@{}lccccc@{}}
\toprule
Model & $A$ & $B$ & $t$ & Seeds $h$ & Exceptions $b$\\
\midrule
Metric & $1.11$ & $1.78$ & $188/125$ & $2$ & $1$\\
Priority & $1$ & $3$ & $1173/500$ & $2$ & $1$\\
Squared metric & $1.45001$ & $3.4032$ & $20543/10000$ & $1$ & $1$\\
\bottomrule
\end{tabular}
\end{center}
The coefficient $t$ multiplies accounted opening cost in the certificate.
For the priority initializer it is three times the growth-process scaling,
because cleanup charges connection costs by a factor of three.
The following sections establish each row: first the metric quota prefix,
then the cleanup needed for priority constraints, and finally the local
certificate needed for squared costs.

\section{Metric facility location: a quota prefix of JMS}\label{sec:metric}
In the metric case, we obtain the initializer by stopping reassigning JMS at
the quota and applying its local cost bound to the optimum's clients. This
section gives the complete transfer at the level of the cost certificate;
Appendices~\ref{sec:algorithm}--\ref{app:implementation} provide the event and
finite-certificate details.

Let $\F,\J$ be finite facility and client sets, with metric connection costs
$c_{ij}$ and quota $p$. Return the empty solution for $p=0$. For $p>0$, use two
seeds, enumerate singleton solutions as well, and set $g_i=0$ on $H$ and
$g_i=\rho f_i$ elsewhere, where $\rho=188/125$.

\paragraph{Growth and truncation.}
Run reassigning JMS on the pruned candidate set. An unconnected client's
budget grows with time. Its budget $\alpha_j$ freezes at first connection;
a connected client's current distance $r_j$ can decrease through reassignment.
A closed facility receives offers $(t-c_{ij})_+$ from unconnected clients
and $(r_j-c_{ij})_+$ from connected clients. When offers pay its fee, open it
and connect or reassign the positive contributors. A client also connects
when the clock reaches its distance to an already open facility. Open
zero-cost facilities at time zero and process tied events in fixed order.

Check the quota after each atomic event. At the first event connecting at
least $p$ clients, remove any excess only from the newly connected batch.
Let $K$ be the retained clients, $S_0$ the open facilities, and $E$ the final
opening facility when deletion was needed, or the empty set otherwise.
Keep $E$ open. Set $w=0$ on $H\cup E$ and $w=f$ elsewhere, then run the common
augmentation procedure. Appendix~\ref{sec:algorithm} specifies all event rules.

\paragraph{A local budget comparison.}
For every facility $i$ and every client subset $V$ of the same complete JMS run,
\begin{equation}\label{eq:metric-local-main}
 \sum_{j\in V}\alpha_j^{\rm full}
 \le1.11g_i+1.78\sum_{j\in V}c_{ij}.
\end{equation}
The arbitrary-star analysis is due to JMS and the coefficients to
MYZ~\cite{JMS,MYZ}; Appendix~\ref{app:local} gives a finite proof with exact
constants. Complete the original prefix without deleting clients or changing
fees. The retained $K$ consists of $p$ smallest frozen budgets in that
continuation. Applying~\eqref{eq:metric-local-main} to the reference optimum's
assignment groups gives
\[
 L:=\sum_{j\in K}\alpha_j\le1.11g(O^\star)+1.78C^\star.
\]

\paragraph{One unpaid opening.}
At each JMS event, opening costs plus current connection costs equal the sum
of connected clients' frozen budgets. If the removed clients form a set $D$
and the final facility is $\ell$, their deleted opening contributions sum to
\[
 \delta=\sum_{j\in D}(T-c_{\ell j})\le g_\ell.
\]
Subtracting their budgets and connections from the event identity gives
$g(S_0)+\sum_{j\in K}r_j=L+\delta$.
Exempting $\ell$ and choosing the cheapest $p$ clients therefore yields
\[
 g(S_0\setminus E)+C_p(S_0)\le L.
\]
The same inequality follows directly when there is no deletion. In the good
seed branch, $g(O^\star)=\rho(F^\star-P)$ and
$g(S_0\setminus E)=\rho w(S_0)$. This is exactly the transfer certificate
with $A=1.11,B=1.78,t=\rho,b=1$.

\begin{theorem}\label{thm:main}
Metric facility location with outliers has a deterministic polynomial-time
$571/376$-approximation serving exactly $p$ clients.
\end{theorem}
\begin{proof}
For $\rho=188/125$, put $a=571/376$. The exact comparisons are
\[
 1.11+\ln\rho<a=1+0.78/\rho,\qquad 3/2<a<1.52.
\]
For example, with $z=63/313$, the positive series for the logarithm gives
$\ln\rho\le2(z+z^3/(3(1-z^2)))=6005349/14711000$;
its gap to $a-1.11$ is $727/1838875>0$.
The transfer theorem applies with $h=2,b=1$.
The full accounting and rational implementation appear in
Appendices~\ref{sec:prefix} and~\ref{app:implementation}.
\end{proof}

\section{Priority facility location: preserving legal connections}\label{sec:priority}
Priority constraints preserve the service-cost structure from
Section~\ref{sec:framework} but change which reassignments are legal. The
initializer must preserve that eligibility through its cleanup.

A candidate is a location--level pair: facility $i$ has level $s_i$, and
client $j$ requires $\ell_j$. An edge is eligible when $s_i\ge\ell_j$;
its cost is the metric distance between the two locations. We use the model
of~\cite{LHSW}, including nondecreasing fees with level at a fixed location.
If several levels are opened at one location, the highest can serve all their
clients, so the output can be consolidated without increasing cost.

The initializer retains contributions after clients freeze. This makes its
price comparison straightforward, but can pay several tentative facilities
with the same money. A cleanup in decreasing service level removes that
overlap and ensures that reassignment remains legal. These two properties
supply the certificate needed by the transfer theorem.

\begin{theorem}\label{thm:priority}
For the priority model above there is a deterministic polynomial-time algorithm with cost at most $a_P\OPT$, where
\[
a_P=\max\left\{1+\ln\frac{1173}{500},\,1+\frac{1000}{1173}\right\}<1.853.
\]
It respects the exact service quota and returns zero cost when $\OPT=0$.
\end{theorem}
\begin{proof}
For $p=0$ return the empty solution. Enumerate all single-facility candidates and all pairs $H$. For a pair put $P=f(H)$, $\tau=\min_{h\in H}f_h$, and retain $H$ and all facilities of original price at most $\tau$. Discard a branch if fewer than $p$ clients have any eligible facility. Set $\rho_P=391/500$, $g=0$ on $H$, and $g_i=\rho_P f_i$ otherwise.

\smallskip\noindent\emph{Growth and exact-quota truncation.}
Open all zero-$g$ facilities at time zero. Unfrozen client prices grow with the common clock; frozen prices stay fixed. On an eligible edge put $\beta_{ij}=(\alpha_j-c_{ij})_+$, retaining this contribution after freezing. Open a closed facility when its total contribution equals its fee, freezing its active positive contributors. Freeze an active client separately when its clock reaches its distance to an open eligible facility. Process simultaneous events one at a time in fixed order and test the quota after each event. At the first attainment of at least $p$, remove excess clients only from the final event's newly frozen clients. Let $K$ be the $p$ retained clients and $E$ the last opening facility if deletion occurred, or the empty set otherwise. Do not change prices or supports after deletion.

\smallskip\noindent\emph{The price comparison.}
For every retained candidate, $\sum_j\beta_{ij}\le g_i$. A closed facility cannot cross its opening threshold. At an opening, all active positive contributors freeze. Subsequent active clients freeze no later than their distance to that open facility; thus its support and contributions never grow again. Zero-cost facilities have no positive contributors. The clients in $K$ have $p$ smallest stopping prices, including ties. Consequently every eligible comparison solution serving $p$ clients at cost $C^r$ satisfies
\[
L:=\sum_{j\in K}\alpha_j\le C^r+g(O).
\]
This is valid even when the comparison clients differ from $K$.

\smallskip\noindent\emph{Cleanup and legal reassignment.}
Let $B_i=\{j:\beta_{ij}>0\}$ for each opened facility, including deleted clients. Scan facilities by nonincreasing service level and retain a facility precisely when its support is disjoint from all previously retained supports. Call the retained set $Q$. A client supports at most one member of $Q$. For $j\in K$ set $(\alpha_j^F,\alpha_j^C)=(\beta_{ij},c_{ij})$ if such $i\in Q$ exists; otherwise use $(0,\alpha_j)$. In both cases the two components sum to $\alpha_j$.

Put $S_0=Q\cup H\cup E\cup\{i:g_i=0\}$ and set $w_i=0$ on $H\cup E$, $w_i=f_i$ otherwise. Every positive contributor of a facility opened before the final event froze before that event and remains in $K$. Thus each paid member of $Q$ is completely funded by $K$. Disjoint supports imply
\[
\rho_P w(S_0)\le\sum_{j\in K}\alpha_j^F.
\]
A positive contributor to $Q$ connects directly at cost $\alpha_j^C$. Otherwise let $a$ witness the first freezing of $j$. Its opening time obeys $t_a\le\alpha_j$ and $c_{aj}\le\alpha_j$. If $a$ was discarded, a retained facility $b$ of level $s_b\ge s_a$ shares a positive contributor $j'$. This contributor froze by $t_a$, so
\[
\alpha_{j'}\le t_a\le\alpha_j,\qquad
c_{aj'},c_{bj'}<\alpha_{j'}.
\]
The metric path $j,a,j',b$ gives $c_{bj}\le3\alpha_j$, and $b$ is eligible for $j$. The level order of cleanup is what guarantees eligibility. If $a\in Q$ connect directly. Therefore
\[
\cp(S_0)\le3\sum_{j\in K}\alpha_j^C.
\]
Combining the two payments and the comparison price bound, for $t=3\rho_P=1173/500$,
\[
tw(S_0)+\cp(S_0)\le t f(O\setminus H)+3C^r
\quad(H\subseteq O).
\]

\smallskip\noindent\emph{Applying the transfer theorem.}
Choose the branch whose seeds are the two most expensive facilities of a fixed optimum. All optimal facilities survive, $P\ge2\tau$, and $R=F^*-P\ge0$. Theorem~\ref{prop:transfer}, with $A=1,B=3$, gives
\[
w(S_g)+\cp(S_g)\le (1+\ln t)R+(1+2/t)C^*\le a_P(\OPT-P).
\]
Every zero-$w$ facility is initially included. The $R=0$ case is included in that theorem.

For $z=673/1673$, the positive logarithmic series gives
\[
\ln t\le 2(z+z^3/3)+\frac{2z^5}{5(1-z^2)}
=U:=\frac{23425470542874593}{27463497192705000}<\frac{853}{1000}.
\]
Also $1000/1173<U$ and $a_P>3/2$. The exception has original price $e\le\tau$, so restoring original fees gives
\[
f(S_g)+\cp(S_g)\le a_P\OPT-(a_P-1)P+e
\le a_P\OPT+(3-2a_P)\tau\le a_P\OPT.
\]
One-facility optima were explicitly enumerated. At any location retain just the highest opened level: this facility was already paid for, and it can serve all clients of the deleted lower levels at the same distance. Nonnegative fees ensure no cost increase.

There are $O(m^2)$ branches, at most $m+n$ growth events and $m$ greedy additions per branch. Between distance breakpoints the growth contribution is affine with integer slope at most $n$. Event times require rational operations and division by such an integer; common denominator bit length grows by at most $O((m+n)\log(n+1))$ beyond input length. Eligibility, cleanup and client selection are polynomial. Thus the statement holds in the rational bit model.
\end{proof}

\section{Squared-metric facility location: certifying the local bound}\label{sec:squared}
The squared-metric application changes a different part of the initializer.
All connections are eligible, so the JMS event accounting and quota truncation
from Section~\ref{sec:metric} apply directly. The finite input matrix now
satisfies the square-root triangle inequality stated in the introduction.
To obtain the joint certificate, we replace the metric local budget bound
by the following inequality.

\begin{lemma}[Squared-metric local certificate]\label{lem:squared-local}
In any complete reassigning JMS run with opening costs $g$, for every candidate
facility $i$ and every client subset $V$,
\[
 \sum_{j\in V}\alpha_j^{\rm full}
 \le A g_i+B\sum_{j\in V}c_{ij},\qquad
 A=1.45001,\quad B=3.4032.
\]
\end{lemma}
\begin{proof}[Proof overview]
Order $V$ by frozen price and put $d_j=c_{ij}$. The same-run constraints are
monotonicity of prices and reassignment distances, a local offer budget, and
\[
 \sqrt{\alpha_l}\le\sqrt{r_{jl}}+\sqrt{d_j}+\sqrt{d_l}\quad(j<l).
\]
These are the squared-metric factor-revealing constraints of~\cite{FMMP}.
For arbitrary $|V|$, replicate every client $k$ times and divide the resulting
sequence into $k$ equal blocks. Averaging preserves the budget by convexity
of the positive-part function and preserves the triangle constraint by
concavity of $(\sqrt{x}+\sqrt{y}+\sqrt{z})^2$. Thus a fixed-size upper-bound
program controls every subset size.

For $k=400$, a rational nonnegative combination of valid rows yields
coefficients $1.450000003915$ and $3.403166095612$, which are bounded by the
stated $A,B$. The rows, their multipliers, and a solver-free verifier are
provided with the paper. Appendix~\ref{app:squared-local} proves the
same-run constraints and finite reduction; Appendix~\ref{app:squared-numeric}
gives the exact coefficient checks, including all rounding residuals.
\end{proof}

\paragraph{From the certificate to a quota.}
Use the metric prefix procedure with one seed $H=\{h\}$, keeping all
facilities of original fee at most $f_h$. Set $g_h=0$ and $g_i=\rho f_i$
otherwise, with $\rho=20543/10000$. Stop at the first quota-reaching event,
remove excess new clients, and exempt its final facility when needed.
The retained-budget comparison and deleted-contribution identity in
Section~\ref{sec:metric} do not use triangle inequalities. With
Lemma~\ref{lem:squared-local}, they give
\[
 \rho w(S_0)+C_p(S_0)\le A\rho(F^\star-P)+BC^\star.
\]
The greedy service-cost analysis is unchanged as well. This proves the
initialization requirements of the transfer theorem with one exception.

\begin{theorem}\label{thm:squared}
Squared-metric facility location with outliers has a deterministic
polynomial-time $(43399/20000)$-approximation serving exactly $p$ clients.
\end{theorem}
\begin{proof}
Let $a=43399/20000=2.16995$ and $z=(\rho-1)/(\rho+1)=10543/30543$.
Exact rational comparison gives
\[
 A+2\sum_{u=0}^{19}\frac{z^{2u+1}}{2u+1}
   +\frac{2z^{41}}{41(1-z^2)}<a,
 \qquad 1+\frac{B-1}{\rho}=\frac{44575}{20543}<a.
\]
The first expression bounds $A+\ln\rho$ from above. Since $a>2$, a single
seed absorbs the one exception: $h(a-1)\ge b$ for $h=b=1$.
Apply Theorem~\ref{prop:transfer}. Its zero-residual case includes an optimum
using only the seed. Event times and density comparisons are rational;
computing square roots and solving the certificate program are not steps of
the approximation algorithm.
\end{proof}

\clearpage
\appendix
\section{Metric algorithm: event rules and cost accounts}\label{sec:algorithm}
Let $\F$ and $\J$ be finite facility and client sets, with
$m=|\F|\ge1$ and $n=|\J|$. Facility $i$ has opening cost $f_i\ge0$.
Facilities and clients lie in a common metric space; $c_{ij}$ denotes their
distance, and distinct input objects may be colocated. All distances are
finite. The integer quota satisfies $0\le p\le n$. For $p=0$, the algorithm
returns the empty solution, so assume $p\ge1$ below.

For nonempty $S\subseteq\F$, define
\begin{equation}\label{eq:service}
 d_j(S)=\min_{i\in S}c_{ij},\qquad
 \Cp(S)=\min_{K\subseteq\J:\,|K|=p}\sum_{j\in K}d_j(S),\qquad
 \Delta_i(S)=\Cp(S)-\Cp(S\cup\{i\}).
\end{equation}
Sorting the distances computes $\Cp(S)$ and a served set attaining it.
Write $v(S)=\sum_{i\in S}v_i$ for any facility-cost vector $v$.
Fix
\begin{equation}\label{eq:parameters}
 \rho=\frac{188}{125},\qquad a=\frac{571}{376}.
\end{equation}
All ties are resolved using a fixed deterministic ordering of facilities,
clients, and event types, independent of the unknown optimum.

\subsection{A prefix of reassigning JMS}
Run the reassigning JMS process~\cite[Algorithm 2]{JMS} on a nonempty candidate
set $I$ with fixed nonnegative opening costs $g$. Let $S$ be the open facilities,
$U$ the connected clients, and $t$ the clock. An unconnected client has budget
$t$. Its budget $\alpha_j$ freezes when it first connects. A connected client
has a current assigned distance $r_j$, which can decrease after reassignment.
Its frozen budget and current distance must be kept distinct.

Initially, open every zero-$g$ facility and set $U=\varnothing$ at time zero.
An unconnected client does not connect to a positive-distance facility merely
because that facility is free. For each closed facility $i$, client $j$ offers
\begin{equation}\label{eq:offers}
 b_{ij}(t)=
 \begin{cases}
  \pos{t-c_{ij}},&j\notin U,\\
  \pos{r_j-c_{ij}},&j\in U.
 \end{cases}
\end{equation}
Advance the clock until an event is possible, and process events as follows.
\begin{enumerate}
\item When $\sum_j b_{ij}(t)=g_i$ for a closed facility, open it and connect
or reassign every client with a strictly positive offer. Newly connected
clients freeze at $\alpha_j=t$; all clients assigned in this event receive
current distance $r_j=c_{ij}$.
\item When an unconnected client satisfies $t=c_{ij}$ for an open facility,
connect that client and freeze its budget at $t$.
\end{enumerate}
Recompute offers and event eligibility after each event. A client with zero
offer at an opening is handled by a single-client event, if needed. Check the
quota immediately after each event, even when other events have the same
timestamp. Stop at the first event with $|U|\ge p$.

Let $T$ be the stopping time and $S_0$ the open set. If $|U|=p$, set $K=U$
and $E=\varnothing$. Otherwise the last event opened a facility $\ell$.
Remove exactly $|U|-p$ clients from that event's newly connected batch, and
call the retained set $K$. Set $E=\{\ell\}$. The facility $\ell$ remains open.

\subsection{Enumeration and augmentation}
For each pair $H\subseteq\F$, set
\begin{equation}\label{eq:pruning}
 P=f(H),\qquad \tau=\min_{h\in H}f_h,\qquad
 I_H=H\cup\{i\in\F:f_i\le\tau\}.
\end{equation}
Run the prefix procedure on $I_H$ with costs $g_i=0$ for $i\in H$ and
$g_i=\rho f_i$ otherwise. After it returns $(S_0,K,E)$, fix residual costs
$w_i=0$ for $i\in H\cup E$ and $w_i=f_i$ otherwise. Every zero-$w$ facility
is already in $S_0$: zero original costs and seeds were opened initially,
and $E\subseteq S_0$.

Starting from $S_0$, repeatedly add a facility in $I_H\setminus S$ of maximum density
$\Delta_i(S)/w_i$ if that density exceeds one. The denominator is positive
for every unopened candidate. At each step, $\Cp$ is recomputed over all
clients. The set $K$ from the prefix proof is not held fixed during augmentation.

\begin{table}[t]
\centering
\begin{tabular}{@{}lll@{}}
\toprule
Cost & Meaning & Definition in a pair branch\\
\midrule
$f_i$ & Original opening cost & Input; used to select the output\\
$g_i$ & JMS opening cost & $0$ on $H$, and $\rho f_i$ elsewhere\\
$w_i$ & Augmentation cost & $0$ on $H\cup E$, and $f_i$ elsewhere\\
\bottomrule
\end{tabular}
\caption{The metric never changes. The exceptional facility is kept open;
only its cost account changes after the prefix is formed.}
\label{tab:accounts}
\end{table}

\paragraph{Algorithm 1: Exact-quota facility location.}
\begin{enumerate}
\item Evaluate each singleton facility set, serving its $p$ cheapest clients.
\item For each pair $H$, form $I_H$ and $g$, run the quota prefix to obtain
$(S_0,K,E)$, and fix $w$. Run maximum-density augmentation from $S_0$,
reselecting clients through $\Cp$, to obtain $S_g$.
\item Return the candidate minimizing $f(S)+\Cp(S)$, with a served set
attaining $\Cp(S)$.
\end{enumerate}
No branch is selected using an optimum or a reference cost. The word
``prepay'' refers to the accounting in the proof; the returned solution
pays every original opening cost.

\section{A certificate for the quota prefix}\label{sec:prefix}
The proof uses a local property of a complete JMS run. Here a \emph{star}
means one candidate facility together with any subset of clients, regardless
of how those clients are assigned by the algorithm.

\begin{lemma}[JMS--MYZ local bound]\label{lem:local}
Complete a JMS run under the event rules above, with costs $g$, and let
$\alpha_j^{\rm full}$ be its frozen budgets. For every candidate facility $i$
and every client subset $V$,
\begin{equation}\label{eq:local}
 \sum_{j\in V}\alpha_j^{\rm full}
 \le \Acoef g_i+\Bcoef\sum_{j\in V}c_{ij}.
\end{equation}
\end{lemma}
The subset is selected for analysis from one run on all clients. JMS's
arbitrary-star construction~\cite[Section 5.1]{JMS} and the coefficients of
MYZ~\cite[Lemma 4]{MYZ} give this interface. Appendix~\ref{app:local} proves
the required statement directly, including equal freeze times, zero costs,
and an explicit finite version of the numerical certificate.

\begin{lemma}[Event accounting]\label{lem:accounting}
After every JMS event,
\begin{equation}\label{eq:accounting}
 g(S)+\sum_{j\in U}r_j=\sum_{j\in U}\alpha_j.
\end{equation}
\end{lemma}
\begin{proof}
The identity holds at time zero. A single-client event adds its connection
distance to both sides. At an opening of $i$, let $N$ be the newly connected
clients and $B$ the reassigned old clients. Tightness gives
\[
 g_i=\sum_{j\in N}(t-c_{ij})+
      \sum_{j\in B}(r_j^--c_{ij}).
\]
The increase in the left side of~\eqref{eq:accounting} is consequently
$g_i+\sum_{j\in N}c_{ij}+\sum_{j\in B}(c_{ij}-r_j^-)=|N|t$.
This is exactly the increase in frozen budgets.
\end{proof}

\begin{lemma}[One exceptional opening]\label{lem:prefix}
Let $(S_0,K,E)$ be any quota prefix. If $O$ is a reference facility set with
an assignment of $p$ clients of connection cost $C^{\rm ref}$ in the same
candidate instance, then
\begin{equation}\label{eq:prefix}
 g(S_0\setminus E)+\Cp(S_0)
 \le \Acoef g(O)+\Bcoef C^{\rm ref}.
\end{equation}
\end{lemma}
\begin{proof}
Continue the original state to a complete JMS run, before removing clients
or changing any costs. Put $L=\sum_{j\in K}\alpha_j$. Every client connected
before time $T$ is retained. Clients removed at time $T$ have budget $T$,
and still-unconnected clients eventually freeze at a time at least $T$.
Thus $K$ is a set of $p$ smallest complete-run budgets. If $K^{\rm ref}$
is the reference client set, Lemma~\ref{lem:local}, applied to its disjoint
assignment groups, gives
\begin{equation}\label{eq:price-comparison}
 L\le\sum_{j\in K^{\rm ref}}\alpha_j^{\rm full}
 \le\Acoef g(O)+\Bcoef C^{\rm ref}.
\end{equation}
Unused reference facilities only add nonnegative cost on the right.

All connection distances in the following accounting refer to the stopping
state, denoted by $r_j(T)$, rather than to the complete continuation. If no
clients are removed, Lemma~\ref{lem:accounting} and
$\Cp(S_0)\le\sum_{j\in K}r_j(T)$ prove the claim.
Otherwise let $D$ be the removed clients and $\ell$ the last facility. Each
$j\in D$ first connects in this event, so its budget is $T$ and its distance
is $c_{\ell j}$. Define its total opening contribution by
\[
 \delta=\sum_{j\in D}(T-c_{\ell j}).
\]
These are a subset of the nonnegative contributions opening $\ell$, including
any savings from old clients. Hence $0\le\delta\le g_\ell$. Removing the
clients' budgets and distances from~\eqref{eq:accounting} gives
\[
 g(S_0)+\sum_{j\in K}r_j(T)=L+\delta.
\]
Keeping $\ell$ open and reselecting the cheapest $p$ clients now yields
\[
 g(S_0\setminus E)+\Cp(S_0)
 \le L+\delta-g_\ell\le L.
\]
Combine this with~\eqref{eq:price-comparison}.
\end{proof}

\subsection{The reference solution in a good branch}
Fix one optimum throughout the analysis, with facilities $O^\star$, served
clients $K^\star$, and opening and connection costs $F^\star,C^\star$.
In particular, $\OPT=F^\star+C^\star$; $C^\star$ need not be the smallest
connection cost among all feasible facility sets. Suppose $|O^\star|\ge2$
and let $H$ be its two most expensive facilities. They define a branch of
the enumeration. Every facility of $O^\star$ remains in $I_H$.

Define the fixed residual \emph{reference} budget and the algorithm's
\emph{initial} costs by
\begin{equation}\label{eq:initial-symbols}
 R=F^\star-P,\qquad F_0=w(S_0),\qquad C_0=\Cp(S_0).
\end{equation}
The quantity $R$ belongs to the reference solution and does not change during
augmentation. In contrast, $F_0$ is the algorithm's cost before augmentation.
Since $g(O^\star)=\rho R$ and $g(S_0\setminus E)=\rho w(S_0)$,
Lemma~\ref{lem:prefix} gives the initialization certificate
\begin{equation}\label{eq:initial-certificate}
 \rho F_0+C_0\le\Acoef\rho R+\Bcoef C^\star.
\end{equation}
Moreover,
\begin{equation}\label{eq:reference-budget}
 w(O^\star)=R-f(E\cap O^\star)\le R.
\end{equation}
The possible inequality is useful: $R$ is an upper bound on the reference
opening cost under $w$, not a definition of that cost.

\section{Greedy augmentation with changing clients}\label{sec:greedy}
Throughout this section the residual costs $w$ and the eligibility relation are fixed. The argument applies to any feasible initial set, including the priority model. All zero-$w$
facilities are in a nonempty initial set $S_0$. Let
$F_0=w(S_0)$ and $C_0=\Cp(S_0)$, and write
$F_g=w(S_g)$ and $C_g=\Cp(S_g)$ for the final costs. In particular,
$F_g-F_0$ is the sum of the additional opening costs, rather than $F_g$ itself.

\subsection{A comparison at the same current set}
\begin{lemma}[Reselection is supermodular]\label{lem:supermodular}
The function $\Cp$ is nonincreasing. For feasible $A\subseteq B$ and
$i\notin B$, we have $\Delta_i(A)\ge\Delta_i(B)$.
\end{lemma}
\begin{proof}
For a cost threshold $u\ge0$, let
$N_u(S)=\{j:d_j(S)\le u\}$ and $H_u(S)=\min\{p,|N_u(S)|\}$.
The cheapest $p$ distances satisfy
\begin{equation}\label{eq:layer-cake}
 \Cp(S)=\int_0^\infty[p-H_u(S)]\,du.
\end{equation}
Indeed, the number of these distances exceeding $u$ is
$p-\min\{p,|N_u(S)|\}$. This remains true at ties, which do not affect the
integral. On supersets of a feasible initial set, the $p$ smallest distances are finite, so the integrand eventually vanishes.

Let $D_u(i)=\{j:i\text{ is eligible for }j,\ c_{ij}\le u\}$. Adding $i$ increases $H_u$ by
\[
 \min\bigl\{|D_u(i)\setminus N_u(S)|,\ (p-|N_u(S)|)_+\bigr\}.
\]
Both arguments of this minimum can only decrease when $S$ grows.
Thus $H_u$ is submodular, and integrating its marginal gains proves the
claimed inequality. Monotonicity follows directly from~\eqref{eq:service}.
\end{proof}
No metric property was used here. The point is that the client set is already
optimized inside $\Cp$, so this argument permits a different served set after
each facility addition.

\begin{corollary}[Reference marginal bound]\label{cor:marginal}
If a facility set $O$ admits a service assignment of $p$ clients with connection
cost at most $C^\star$, then, for every feasible $S$,
\begin{equation}\label{eq:marginal}
 \sum_{i\in O\setminus S}\Delta_i(S)
 \ge\Cp(S)-\Cp(S\cup O)\ge\Cp(S)-C^\star.
\end{equation}
\end{corollary}
\begin{proof}
Add the facilities of $O\setminus S$ one at a time. Their successive savings
telescope to $\Cp(S)-\Cp(S\cup O)$. Each is at most the corresponding saving
measured at the original set $S$, by Lemma~\ref{lem:supermodular}.
The union can use the reference assignment, giving the second inequality.
\end{proof}

\subsection{The discrete threshold crossing}
Assume $w(O)\le R$ for the reference set in Corollary~\ref{cor:marginal}.
For now suppose $R>0$. If the current connection cost $C$ exceeds
$C^\star+R$, some unopened reference facility has density at least
$(C-C^\star)/R>1$. To see this, combine~\eqref{eq:marginal} with
$w(O\setminus S)\le R$ and use a weighted average; every unopened facility
has positive cost. Therefore greedy cannot stop above this threshold.
If the chosen facility costs $v$ and the new connection cost is $C'$, then
\begin{equation}\label{eq:one-step}
 v\le R\frac{C-C'}{C-C^\star}.
\end{equation}

\begin{lemma}[Cumulative augmentation cost]\label{lem:log}
If $R>0$ and $C_0>C^\star+R$, then
\begin{equation}\label{eq:log-bound}
 F_g+C_g\le F_0+C^\star+R+
 R\ln\frac{C_0-C^\star}{R}.
\end{equation}
If $C_0\le C^\star+R$, then $F_g+C_g\le F_0+C_0$.
\end{lemma}
\begin{proof}
Write $D=C-C^\star$. For a step staying above the threshold, with
$D>D'>R$, equation~\eqref{eq:one-step} gives
\[
 v\le R\frac{D-D'}{D}\le\int_{D'}^D\frac{R}{x}\,dx.
\]
There are finitely many additions, and the algorithm cannot stop with $D>R$.
Thus, when $D_0=C_0-C^\star>R$, a first crossing exists. Denote the gap just
before it by $s>R$, and just after it by $d\le R$. Integrating over all
preceding steps bounds their total opening cost by $R\ln(D_0/s)$.

The crossing step needs a separate bound because $d$ may be zero or negative.
Using~\eqref{eq:one-step} and $1-R/s>0$, its opening cost satisfies
\begin{align*}
 v+d&\le R+d(1-R/s)\\
     &\le R+R(1-R/s)\\
     &\le R+R\ln(s/R).
\end{align*}
We combine $v$ with $d$, not with the full remaining connection cost.
The latter is $C^\star+d$. At the crossing, total cost is therefore at most
\[
 F_0+R\ln(D_0/s)+v+C^\star+d
 \le F_0+C^\star+R+R\ln(D_0/R).
\]
Every subsequent accepted addition costs less than its connection saving,
so total residual cost strictly decreases. The same monotonicity proves the
second case when the initial state is already below the threshold.
\end{proof}

\subsection{Using the initialization certificate}
The next lemma explains why the opening and connection costs of the initial
solution should be bounded together. A larger $C_0$ leaves a smaller allowable
$F_0$, and this compensates for a longer augmentation phase.

\begin{lemma}[From initialization to final cost]\label{lem:bifactor}
Suppose $A,B\ge1$, $\rho\ge1$, and
\begin{equation}\label{eq:generic-certificate}
 \rho F_0+C_0\le A\rho R+B C^\star.
\end{equation}
With the reference assumptions above, including $w(O)\le R$, greedy returns
\begin{equation}\label{eq:bifactor}
 F_g+C_g\le(A+\ln\rho)R+
 \left(1+\frac{B-1}{\rho}\right)C^\star.
\end{equation}
This includes $R=0$.
\end{lemma}
\begin{proof}
First let $R>0$ and $x=(C_0-C^\star)/R>1$. The certificate gives
$F_0\le(A-x/\rho)R+(B-1)C^\star/\rho$.
Substituting into Lemma~\ref{lem:log} yields
\[
 F_g+C_g\le(A+1+\ln x-x/\rho)R+
 \left(1+\frac{B-1}{\rho}\right)C^\star.
\]
The function $1+\ln x-x/\rho$ is maximized at $x=\rho$, where its value is
$\ln\rho$. This proves~\eqref{eq:bifactor} in the first case.

If $R>0$ but $C_0\le C^\star+R$, monotonicity of total cost gives
\begin{align*}
 F_g+C_g&\le F_0+C_0\\
 &\le AR+\frac{B}{\rho}C^\star+(1-1/\rho)C_0\\
 &\le(A+1-1/\rho)R+
       \left(1+\frac{B-1}{\rho}\right)C^\star.
\end{align*}
Use $1-1/\rho\le\ln\rho$.

Finally, if $R=0$, then $w(O)=0$ and hence $O\subseteq S_0$.
Thus $C_0\le C^\star$. The same first two lines above imply
\[
 F_g+C_g\le\frac{B}{\rho}C^\star+(1-1/\rho)C_0
 \le\left(1+\frac{B-1}{\rho}\right)C^\star.
\]
This case uses neither division by $R$ nor a logarithm of a gap.
\end{proof}

\section{Restoring costs and proving the approximation}\label{sec:restoration}
Return to the fixed optimum and its good pair branch. Equations
\eqref{eq:initial-certificate} and~\eqref{eq:reference-budget} supply the
hypotheses of Lemma~\ref{lem:bifactor} with $A=1.11$ and $B=1.78$.
Our rational parameters satisfy
\begin{equation}\label{eq:parameter-inequalities}
 1.11+\ln\rho<a=1+\frac{0.78}{\rho},\qquad
 \frac32<a<\frac{38}{25}.
\end{equation}
For completeness, set $z=(\rho-1)/(\rho+1)=63/313$. The logarithm series gives
\[
 \ln\rho\le2\left(z+\frac{z^3}{3(1-z^2)}\right)
 =\frac{6005349}{14711000},\qquad
 a-\frac{111}{100}-\frac{6005349}{14711000}
 =\frac{727}{1838875}>0.
\]
The other comparisons in~\eqref{eq:parameter-inequalities} are rational.
It follows that
\begin{equation}\label{eq:residual-final}
 w(S_g)+\Cp(S_g)\le a(R+C^\star)=a(\OPT-P).
\end{equation}

Both $H$ and $E$ remain open. If $E=\{\ell\}$, then $\ell\notin H$,
because the seeds were opened initially, whereas $\ell$ opened at the final
overflow event. Pruning guarantees $f_\ell\le\tau$. In either case put
$e=f(E)\le\tau$. The original cost is exactly
\begin{equation}\label{eq:exact-restoration}
 f(S_g)+\Cp(S_g)=w(S_g)+\Cp(S_g)+P+e.
\end{equation}
Combining~\eqref{eq:residual-final}--\eqref{eq:exact-restoration} and using
$P\ge2\tau$ gives
\begin{align*}
 f(S_g)+\Cp(S_g)
 &\le a(\OPT-P)+P+\tau\\
 &=a\OPT-(a-1)P+\tau\\
 &\le a\OPT+(3-2a)\tau\le a\OPT.
\end{align*}
The sign of $-(a-1)$ is essential: prepaying two reference facilities creates
enough slack to pay for the one exceptional facility. If an optimum uses one
facility, its singleton candidate is already evaluated. Every candidate
serves exactly $p$ clients, and selection uses original costs. If $\OPT=0$,
nonnegativity and the same upper bound force the output cost to be zero.

There are $O(m^2)$ pair branches. Each contains at most $m$ opening events,
$n$ first connections, and $m$ greedy additions. Event times are obtained
from piecewise-linear equations with slopes at most $n$; all current
connection costs are input distances. Appendix~\ref{app:implementation}
gives the rational bit-complexity argument. This completes the proof of
Theorem~\ref{thm:main}.

\section{A finite proof of the local JMS--MYZ bound}\label{app:local}
We prove Lemma~\ref{lem:local} on the precise class of arrays obtained from
JMS trajectories. The factor-revealing constraints and multiplier construction
are due to JMS and MYZ~\cite{JMS,MYZ}. We use the parameters in the 2006
version of MYZ, give the finite-size error explicitly, and justify replication
for the trajectory class. This appendix is independent of a numerical LP solver.

\subsection{From one run to an arbitrary star}
Fix a candidate facility and a nonempty subset of $k$ clients. Write $f$ for
its opening cost in this run and $d_j$ for its distance to client $j$.
Order the subset so that $\alpha_1\le\cdots\le\alpha_k$, breaking ties
arbitrarily. For $j<i$, let $r_{j,i}$ be client $j$'s current connection
distance just before clock time $\alpha_i$ if $\alpha_j<\alpha_i$;
if $\alpha_j=\alpha_i$, put $r_{j,i}=\alpha_j$. These are auxiliary values
for the star analysis, not new states of the algorithm.

The following constraints hold:
\begin{align}
 &0\le\alpha_1\le\cdots\le\alpha_k,\quad d_j,f,r_{j,i}\ge0,\label{eq:lp-order}\\
 &r_{j,i}\ge r_{j,i+1} &&(j<i<k),\label{eq:lp-monotone}\\
 &\alpha_i\le r_{j,i}+d_j+d_i &&(j<i),\label{eq:lp-triangle}\\
 &\sum_{j<i}\pos{r_{j,i}-d_j}
   +\sum_{j\ge i}\pos{\alpha_i-d_j}\le f &&(1\le i\le k),\label{eq:lp-budget}\\
 &r_{j,i}\le\alpha_j &&(j<i).\label{eq:lp-trajectory}
\end{align}
Current distances only decrease and never exceed the first connection budget;
this proves~\eqref{eq:lp-monotone} and~\eqref{eq:lp-trajectory}, including
the transition from equal to later freeze times.
For~\eqref{eq:lp-triangle}, if $j$ is already assigned to an open facility
$h$, then client $i$, still unconnected just before $\alpha_i$, has
$\alpha_i\le c_{hi}\le c_{hj}+d_j+d_i$ by the metric inequality.
At a tied freeze time the inequality holds because $r_{j,i}=\alpha_i$.

If the reference facility is still closed just before $\alpha_i$,
the left side of~\eqref{eq:lp-budget} is its offer from this subset in that
left-limit state. It is at most the offer from all clients, which cannot
exceed $f$ before an opening event. All clients with freeze time $\alpha_i$
are unconnected in this left-limit state. If the facility opened strictly
earlier, already connected clients have current distance at most $d_j$,
and still-unconnected clients have clock at most $d_j$. Every term is then
zero. Zero-cost facilities fall into this second case for positive clock
times; at time zero every term is zero directly. This proves the budget row.
An empty subset satisfies the desired bound trivially.

We will prove that every finite array satisfying
\eqref{eq:lp-order}--\eqref{eq:lp-trajectory} obeys
\begin{equation}\label{eq:finite-target}
 \sum_{j=1}^k\alpha_j\le\Acoef f+\Bcoef\sum_{j=1}^k d_j.
\end{equation}
The additional trajectory condition~\eqref{eq:lp-trajectory} is used only
when reducing a small array to a large one.

\subsection{Weighted budget rows}
First consider an array of size $N\ge200000$, with $100000$ dividing $N$.
Set
\[
 p_1=\frac{225}{1000},\quad p_2=\frac{791}{1000},\quad
 p_3=\frac{30499}{100000},\quad p_4=\frac{6984}{100000},
\]
and define integer breakpoints $k_r=p_rN$ for $r=1,2,3$, together with
$t=N-k_3$, $h=N-k_2$, and $q=p_4N$. Define
\[
 s_i=\max\bigl\{0,\max_{\ell\ge i}(\alpha_\ell-d_\ell)\bigr\}.
\]
Then $s_i$ is nonnegative and nonincreasing, $\alpha_i\le s_i+d_i$, and
$r_{j,i}\ge s_i-d_j$ for $j<i$. The latter follows by combining
\eqref{eq:lp-monotone} with~\eqref{eq:lp-triangle} for every $\ell\ge i$;
the extra candidate zero is valid because $r_{j,i}\ge0$.

For any $\ell\ge i$, the $i$th budget row implies
\begin{equation}\label{eq:weighted-row}
 (\ell-i+1)\alpha_i
 \le f+\sum_{j=i}^{\ell}d_j-\sum_{j<i}\pos{r_{j,i}-d_j}.
\end{equation}
Use the following choices of $\ell$ and row multipliers:
\begin{center}
\begin{tabular}{@{}lll@{}}
\toprule
Row indices & $\ell$ & Multiplier\\
\midrule
$1\le i\le k_1$ & $k_2$ & $1/(k_2-i+1)$\\
$k_1<i\le k_3$ & $N$ & $1/(N-i+1)$\\
$k_3<i\le t$ & $N$ & $1/t$\\
$t<i\le k_2$ & $N$ & $q/((N-i)(N-i+1))$\\
\bottomrule
\end{tabular}
\end{center}
The first two groups have left coefficient one. In the third and fourth
groups, complete that coefficient to one using $\alpha_i\le s_i+d_i$,
with respective missing coefficients $(i-k_3-1)/t$ and $1-q/(N-i)$.
Both are nonnegative. For $i>k_2$, use
$\alpha_i\le s_{k_2+1}+d_i$.

To bound the negative terms in~\eqref{eq:weighted-row}, use
\[
 \pos{r_{j,i}-d_j}\ge\pos{s_i-2d_j}
 \ge \tfrac12s_i-d_j
\]
in the first group, and $\pos{r_{j,i}-d_j}\ge s_i-2d_j$ in the next two.
In the fourth group retain the stronger bound
$\pos{r_{j,i}-d_j}\ge\pos{s_{k_2+1}-2d_j}$.

Write $H_b=\sum_{i=1}^b1/i$ and $H_0=0$, and abbreviate
\begin{equation}\label{eq:harmonic-quantities}
 \begin{split}
 U&=H_{k_2}-H_{k_2-k_1},\qquad V=H_{N-k_1}-H_t,\\
 W&=(t-k_3)/t,\qquad Z=q(1/h-1/k_3),\qquad
 \zeta=U+V+W+Z.
 \end{split}
\end{equation}
In the fourth group, drop negative terms for $j>t$. For $k_3<j\le t$,
replace the retained positive part by $s_{k_2+1}-2d_j$.
Adding the weighted rows gives
\begin{equation}\label{eq:raw-coefficients}
 \begin{split}
 \sum_i\alpha_i-\zeta f
 \le{}&\sum_j\lambda_j^{(0)}d_j+\sum_{j=1}^{k_2}\mu_j s_j
       +[h-(t-k_3)Z]s_{k_2+1}\\
     &{}-Z\sum_{j=1}^{k_3}\pos{s_{k_2+1}-2d_j},
 \end{split}
\end{equation}
where direct finite summation yields
\begin{equation}\label{eq:raw-lambda}
 \lambda_j^{(0)}=
 \begin{cases}
 U+2V+2W,&j\le k_1,\\
 U+V+H_{N-j}-H_t+2W,&k_1<j\le k_3,\\
 U+V+2W+2Z-1/t,&k_3<j\le t,\\
 U+V+W+1-q/k_3,&t<j\le k_2,\\
 V+W+1+Z,&j>k_2,
 \end{cases}
\end{equation}
and
\begin{equation}\label{eq:mu}
 \mu_j=
 \begin{cases}
 -(j-1)/(2(k_2-j+1)),&j\le k_1,\\
 -(j-1)/(N-j+1),&k_1<j\le k_3,\\
 -k_3/t,&k_3<j\le t,\\
 1-q/(N-j),&t<j\le k_2.
 \end{cases}
\end{equation}
For example, the first group's distance coefficient for every $j\le k_2$
is $U$: rows $i\le j$ contribute from their distance sum, and rows $i>j$
contribute from the relaxed negative offer. The other coefficients follow
by the same separation of these two contributions.

Raise the third distance coefficient by $1/t$, which is valid because
$d_j\ge0$, and put $\lambda_T=U+V+2W+2Z$.
The bounds proved below give $V\le2Z$. Therefore, for $j\le k_3$,
\[
 \delta_j=\frac{\lambda_T-\lambda_j^{(0)}}{2Z}\in[0,1].
\]
Apply $\pos{x-2d_j}\ge\delta_j(x-2d_j)$ in the last term of
\eqref{eq:raw-coefficients}. Every distance coefficient for $j\le t$
becomes $\lambda_T$. The tail coefficient becomes
\begin{equation}\label{eq:mu-tail}
 \mu_{k_2+1}=h-tZ+\frac{NV-k_3+k_1}{2}.
\end{equation}
Here the finite identity
$\sum_{j=k_1+1}^{k_3}(H_{N-j}-H_t)=(N-k_1)V-(k_3-k_1)$
accounts for the second group. We have established
\begin{equation}\label{eq:certificate-form}
 \sum_i\alpha_i\le\zeta f+\sum_i\lambda_i d_i+
                   \sum_{i=1}^{k_2+1}\mu_i s_i,
\end{equation}
where $\lambda_i=\lambda_T$ for $i\le t$, and the last two cases
of~\eqref{eq:raw-lambda} give the remaining $\lambda_i$.

\subsection{Nonpositive prefix sums and exact constants}
Let $M_b=\sum_{i=1}^b\mu_i$. Equation~\eqref{eq:mu} is nonpositive up to
$t$, and nonnegative from $t+1$ to $k_2$, since $q<h$.
It is enough to check $M_{k_2}\le0$ and $M_{k_2+1}\le0$.
Summing~\eqref{eq:mu}--\eqref{eq:mu-tail} gives
\begin{align}
 M_{k_2}={}&-\tfrac12 k_2U+\tfrac12 k_1-NV+k_3-k_1-k_3W
          +k_2-t-q(H_{k_3-1}-H_{h-1}),\label{eq:prefix-b}\\
 M_{k_2+1}={}&\tfrac32 k_3-k_3W-tZ-\tfrac12 k_2U-\tfrac12 NV
          -q(H_{k_3-1}-H_{h-1}).\label{eq:prefix-end}
\end{align}
In particular, the last harmonic difference has endpoints $k_3-1$ and
$h-1$, as required by summing $1/(N-i)$ for $t<i\le k_2$.

Define the corresponding continuous quantities
\[
 u=\ln\frac{p_2}{p_2-p_1},\quad
 v=\ln\frac{1-p_1}{1-p_3},\quad
 w=\frac{1-2p_3}{1-p_3},\quad
 z=\frac{p_4}{1-p_2}-\frac{p_4}{p_3},\quad
 L=\ln\frac{p_3}{1-p_2},
\]
and
\begin{align*}
 T_b&=-\tfrac12p_1+p_2+2p_3-1-\tfrac12p_2u-v-p_3w-p_4L,\\
 T_e&=-\tfrac12p_2u-\tfrac12v+\tfrac32p_3-p_3w-(1-p_3)z-p_4L.
\end{align*}
For integers $n>m\ge1$, integration of $1/x$ gives
\begin{equation}\label{eq:harmonic-error}
 0\le\ln(n/m)-(H_n-H_m)\le1/m-1/n,
 \qquad H_{n-1}-H_{m-1}\ge\ln(n/m).
\end{equation}
Consequently $U\le u$, $V\le v$, $W=w$, $Z=z$, and
\begin{align}
 M_{k_2}&\le NT_b+E_b,&
 E_b&=\frac{p_1}{2(p_2-p_1)}+
       \frac{p_3-p_1}{(1-p_3)(1-p_1)}<0.35,\label{eq:finite-b}\\
 M_{k_2+1}&\le NT_e+E_e,&
 E_e&=\frac{p_1}{2(p_2-p_1)}+
       \frac{p_3-p_1}{2(1-p_3)(1-p_1)}<0.274.\label{eq:finite-end}
\end{align}

Table~\ref{tab:certificate} records rational interval bounds, with decimal
endpoints interpreted exactly. They are certified by the following finite
formula. For each rational $x>1$, set $y=(x-1)/(x+1)$ and use
\begin{equation}\label{eq:log-certificate}
 L_{48}(x)=2\sum_{j=0}^{47}\frac{y^{2j+1}}{2j+1},\qquad
 L_{48}(x)\le\ln x\le
 L_{48}(x)+\frac{2y^{97}}{97(1-y^2)}.
\end{equation}
Every entry follows by rational arithmetic and outward interval propagation.
Thus the numerical signs do not depend on floating-point rounding.

\begin{table}[t]
\centering
\begin{tabular}{@{}lrr@{}}
\toprule
Expression & Lower endpoint & Upper endpoint\\
\midrule
$v-2z$ & $-0.101406334524$ & $-0.101406334523$\\
$u+v+w+z$ & $1.109984031631$ & $1.109984031632$\\
$u+v+2w+2z$ & $1.776327378295$ & $1.776327378296$\\
$u+v+w+1-p_4/p_3$ & $1.775821352205$ & $1.775821352206$\\
$v+w+1+z$ & $1.775280142066$ & $1.775280142067$\\
$T_b$ & $-0.150379625875$ & $-0.150379625874$\\
$T_e$ & $-0.000001517538$ & $-0.000001517537$\\
\bottomrule
\end{tabular}
\caption{Exact enclosures for the seven conditions in the finite certificate.}
\label{tab:certificate}
\end{table}

For every allowed $N\ge200000$, the upper bounds in
\eqref{eq:finite-b} and~\eqref{eq:finite-end} are negative; in particular,
\[
200000(-0.000001517537)+0.274<0.
\]
All prefix sums $M_b$ are therefore nonpositive. Summation by parts gives
\[
 \sum_{i=1}^{k_2+1}\mu_i s_i
 =\sum_{i=1}^{k_2}M_i(s_i-s_{i+1})+M_{k_2+1}s_{k_2+1}\le0.
\]
The same table gives $V\le v<2z=2Z$, $\zeta<1.11$, and
$\lambda_i<1.7764<1.78$ for every $i$.
Equation~\eqref{eq:certificate-form} proves~\eqref{eq:finite-target}
for these large arrays, including zero distances and zero opening cost.

\subsection{All finite star sizes}
Take any $k$-client array satisfying
\eqref{eq:lp-order}--\eqref{eq:lp-trajectory}. Replace each client by
$Q=200000$ consecutive copies, keep its $\alpha$ and $d$ values, and
replace $f$ by $Qf$. For two copies of the same original client $j$, set
the auxiliary $r'$ value to $\alpha_j$; between copies of different original
clients $j<i$, set it to $r_{j,i}$.

The order and triangle constraints persist. Along a row, the transition
from its own copy block to a later block is nonincreasing precisely because
$r_{j,i}\le\alpha_j$; later transitions use~\eqref{eq:lp-monotone}.
For a budget row inside the copy block of client $i$, its earlier copies
and its not-yet-counted copies together contribute exactly
$Q\pos{\alpha_i-d_i}$. Every other block contributes $Q$ times its
original contribution. The row is therefore at most $Qf$.
The enlarged array has $N=Qk\ge200000$ and $100000\mid N$.
Applying the large-array bound and dividing by $Q$ proves
\eqref{eq:finite-target} for every finite $k$, and completes
Lemma~\ref{lem:local}. The copies exist only in this proof; the algorithm
never constructs them.

\section{Implementation and rational bit complexity}\label{app:implementation}
We spell out why the event description defines a polynomial-time algorithm.
Assume the input costs and distances are nonnegative rationals in binary.
Zero-cost facilities are opened at initialization. At any later state, offers
from connected clients are fixed until an event, while an unconnected client's
offer to $i$ is the piecewise-linear function $\pos{t-c_{ij}}$.
Sorting its distance breakpoints permits exact computation of the next
tightening time for each closed facility. Single-client events are found
from the distances to already open facilities. The earliest eligible event
is processed according to the fixed tie order, after which offers are recomputed.

Between events every relevant offer function is continuous and nondecreasing.
An opening or connection does not increase any other facility's offers:
a newly connected client replaces budget $t$ by a distance at most $t$,
and a reassigned client's distance decreases. Consequently no closed facility
acquires an unprocessed offer strictly larger than its opening cost.
There are at most $m$ openings and $n$ first connections in a complete run;
reassignments occur within openings. As long as a client remains unconnected,
a future connection or opening event exists, since the candidate set is
nonempty and all distances and costs are finite.

For the bit bound, let $D$ be $125$ times the product of all input
denominators. It is a common denominator for all input distances and all
scaled costs $g_i$, including the factor $\rho=188/125$.
Its bit length is at most the sum of the input denominator lengths plus a
constant. Every current connection distance is an input distance. On a
linear piece, a tightening time solves $bt=a/D$ with integer
$1\le b\le n$ and an integer $a$ of polynomial bit length. Thus event
times have polynomial bit length; denominators do not recursively accumulate
previous event denominators. The other event times are input distances.
Exact comparison also resolves simultaneous events and zero offers.

There are $O(m^2)$ pair branches and $m$ singleton candidates. Greedy
augmentation adds at most $m$ facilities per branch. For any current set,
each candidate marginal is obtained by updating the $n$ nearest-facility
distances and sorting them to sum the cheapest $p$. Density comparisons can
be performed by cross multiplication. All sums, comparisons, and branch
evaluations have polynomial bit complexity. This proves the running-time
claim without an oracle for continuous optimization or logarithms.

\section{Same-run local interface for squared costs}\label{app:squared-local}
Fix a full A2/JMS run, a comparison facility $h$, and any nonempty client subset, sorted by its frozen prices. Write $d_j=c_{hj}$ and $f=g_h$. For equal frozen prices set $r_{jl}=\alpha_j$; otherwise use the connection cost of $j$ just before the entire timestamp $\alpha_l$. Then
\begin{align}
\alpha_l&\le\alpha_{l+1}, &0\le r_{jl}&\le\alpha_j,&r_{jl}&\ge r_{j,l+1},\label{ord}\\
\sqrt{\alpha_l}&\le\sqrt{r_{jl}}+\sqrt{d_j}+\sqrt{d_l},&&&&\label{tri}\\
\sum_{j<l}(r_{jl}-d_j)_++\sum_{j\ge l}(\alpha_l-d_j)_+&\le f.&&&&\label{bud}
\end{align}
To prove \eqref{tri} for unequal prices, $j$ is connected to an already open facility $a$, while $l$ remains active; hence $\alpha_l\le c_{al}$ and the squared three-hop inequality applies through $j,h$. Equal prices are immediate. For \eqref{bud}, unopened facilities receive offers at most their fees; restricting to a client subset only drops nonnegative terms. If $h$ is open, all these positive parts vanish. Equal-price auxiliary variables reproduce the left-limit offers, not post-event distances.

\begin{lemma}[Finite transfer to a fixed-dimensional certificate]
Fix $k\ge2$. Let $\mathrm{UP}(k)$ consist of \eqref{ord}, \eqref{tri}, nonnegativity, and the budgets
\[
\sum_{a<b}(r_{ab}-d_a)_++\sum_{a>b}(\alpha_b-d_a)_+\le f.
\]
If every point in this system satisfies $\sum_{b=1}^k\alpha_b\le Af+B\sum_{b=1}^kd_b$, then the same local inequality holds for every finite subset of every full A2 trajectory.
\end{lemma}
\begin{proof}
For a subset of size $n$, replicate each client $k$ times consecutively, keeping its $\alpha,d$ and multiplying $f$ by $k$. Between copies of the same client put $r=\alpha$; between distinct originals keep the original $r$. The cap $r_{jl}\le\alpha_j$ ensures monotonicity at a replication boundary. Every budget is exactly $k$ times the original budget.

Partition the resulting $kn$ entries into $k$ consecutive blocks of size $n$. Average $\alpha,d$ over their blocks, $r$ over each pair of blocks, and divide the replicated $f$ by $n$. The ordering constraints and cap survive averaging. The function
\[
G(x,y,z)=x+y+z+2\sqrt{xy}+2\sqrt{xz}+2\sqrt{yz}
\]
is concave on the nonnegative orthant, so Jensen preserves \eqref{tri}. For each block, omit all within-block terms in \eqref{bud}, average the remaining rows and divide by $n$. Convexity of the positive-part function yields precisely the displayed UP budget. Every objective total has been multiplied by the same $k/n$, proving the claim. If all $d$ vanish in the original $n$-client system, \eqref{tri} and its last budget directly give $f\ge n\alpha_n\ge\sum\alpha$; no normalization by zero is needed.
\end{proof}

\section{Exact finite-certificate protocol}\label{app:squared-numeric}
For positive integers $u,v,w$ and $q=u+v+w$, squared three-hop constraints imply the rational row
\[
\alpha_l-\frac qu r_{jl}-\frac qv d_l-\frac qw d_j\le0
\]
by weighted Cauchy--Schwarz. Other valid rows are ordering, caps, and any subset expansion of a UP budget. More generally, for $0\le e_j\le h$, a weighted budget row is valid:
\[
\sum_{j<l}e_j(r_{jl}-d_j)+\sum_{j>l}e_j(\alpha_l-d_j)-hf\le0.
\]
Indeed each signed term is at most its positive part, and their weighted sum is at most $h$ times the full positive-part budget. A finite list of nonnegative rational row multipliers is summed using exact fraction arithmetic.

To repair rounding deficits, write $D=\sum d_j$. The full UP constraints imply $r_{jl}\le f+d_j\le f+D$, $\alpha_1\le f+d_2$, and, for $l>1$,
\[
\alpha_l\le3r_{1l}+3d_1+3d_l\le3f+6D.
\]
Hence all $\alpha$ and $r$ variables are bounded by $3f+6D$. A column deficit $\epsilon$ can be paid by the valid row $\epsilon(v-3f-6D)\le0$. After adding every such row, require all $\alpha$ coefficients to be at least one and all $r$ coefficients nonnegative. The exact negative facility coefficient and the largest negative distance coefficient give certified $A,B$. No solver tolerance is discarded. The independent verifier rebuilds the rows, rather than trusting a stored solver status or objective.

\begin{lemma}[Independently certified numerical inequality]
Every feasible point of $\mathrm{UP}(400)$ satisfies
\[
\sum_{j=1}^{400}\alpha_j\le
\frac{290000000783}{200000000000}f+
\frac{850791523903}{250000000000}\sum_{j=1}^{400}d_j.
\]
Consequently the same-run local certificate holds with the simpler constants
$A=145001/100000=1.45001$ and $B=2127/625=3.4032$.
\end{lemma}
\begin{proof}
The supplementary file \path{square-k400-certificate.json.gz} contains
123414 nonnegative rational weighted rows: 363 $\alpha$ monotonicity rows,
75746 $r$ monotonicity rows, 46954 weighted-Cauchy rows, and 351 weighted
budget rows. The independent standard-library verifier reconstructs every row
and checks all 80601 columns. There are no cap rows. Upward-rounded repairs
on 342 $\alpha$ columns have total weight $261/200000000000$; there are no
$r$ deficits. Adding the valid residual-payment rows described above and
rounding the distance coefficient upward gives precisely the displayed
inequality. Thus no floating tolerance or solver objective is used as proof.
The finite-transfer lemma then applies to every client-subset size.
\end{proof}

\paragraph{Proof data.}
The accompanying supplementary files contain the certificate,
\path{verify_square_certificate.py}, and instructions for replaying the
exact check using only the Python standard library. The certificate records
rational row multipliers; the verifier reconstructs and sums the rows and
pays all residual column deficits by the bounds proved above. These files
are part of the finite numerical proof of Lemma~\ref{lem:squared-local}.
The mathematical derivation of each allowed row family is given here.

\end{document}